\documentclass[%
aip,
amsmath,amssymb,
reprint,%
]{revtex4-1}

\usepackage{graphicx}
\usepackage{dcolumn}
\usepackage{bm}
\usepackage{physics}
\usepackage{mathrsfs}

\usepackage[utf8]{inputenc}
\usepackage[T1]{fontenc}
\usepackage{mathptmx}

\usepackage{float}
\usepackage[caption=false]{subfig}

\usepackage{hyperref}
\hypersetup{colorlinks=true, citecolor=blue, urlcolor=blue, linkcolor=blue}

\newcommand{\avg}[1]{\mbox{$\langle{#1}\rangle$}}

\begin{document}
    \title{Statistical analysis of electron bunch longitudinal profile reconstructions using the Gerchberg-Saxton algorithm}
    
    \author{Bricker Ostler}
    \email{bostler@lanl.gov}
    \author{Nikolai Yampolsky}
    \author{Quinn Marksteiner}
    \affiliation{Los Alamos National Laboratory, Los Alamos, New Mexico 87545, USA}
    \date{\today}
    
    \begin{abstract}
        Knowledge of longitudinal electron bunch profiles is vital to optimize the performance of plasma wakefield accelerators and x-ray free electron laser linacs. Because of their importance to these novel applications, noninvasive frequency domain techniques are often employed to reconstruct longitudinal bunch profiles from coherent synchrotron, transition, or undulator radiation measurements. In this paper, we detail several common reconstruction techniques involving the Kramers-Kronig phase relationship and Gerchberg-Saxton algorithm. Through statistical analysis, we draw general conclusions about the accuracy of these reconstruction techniques and the most suitable candidate for longitudinal bunch reconstruction from spectroscopic data.
    \end{abstract}
    
    \maketitle
    
    \section{Introduction}
    
    Ultrashort electron bunches are routinely achieved in high-gradient plasma wakefield accelerators and x-ray free electron laser linacs. With these novel applications comes a need for diagnostics that can nondestructively resolve the longitudinal profile of these bunches with femtosecond (fs) resolution in a single shot. The two primary time domain techniques to determine longitudinal electron bunch profiles employ transverse deflecting structures or electro-optic coupling methods, but both techniques are limited in their ability to characterize bunch profiles nondestructively, with fs resolution, and in a single shot. Specifically, transverse deflecting structures can achieve approximately $10$ fs resolution in a single shot but destroy the bunch when sweeping it transversely.\cite{Ego15} Electro-optic techniques, which use the change in birefringence of an electro-optic crystal to encode the bunch profile into a spectral of temporal modulation of a laser pulse, allow for nondestructive and single-shot diagnostics to determine bunch profiles. However, the best attainable resolution of electro-optic techniques is approximately $60$ to $70$ fs.\cite{Berden07, Wilke02}
    
    Longitudinal bunch profiles can be indirectly recovered by detecting coherent synchrotron, transition, or undulator radiation over a large range of frequencies. The intensity distribution $I(\omega)$ of coherent radiation emitted by an electron bunch is directly related to the electron bunch form factor
    
    \begin{equation}
		F(\omega) = \int_{-\infty}^{\infty} f(t) e^{-i\omega t} \dd{t} ,
    \end{equation}
    
    \noindent through $I(\omega) \propto \abs{F(\omega)}^2$. Here, the form factor $F(\omega)$ is defined as the Fourier transform of the normalized longitudinal bunch profile $f(t)$, proportional to the beam current.\cite{Grimm06, Kim17} By detecting a large portion of the intensity distribution of coherent radiation emitted by an electron bunch, the longitudinal bunch profile can be recovered. This allows for single-shot and nondestructive determination of longitudinal bunch profiles with no fundamental limit in resolution. 
    
    Although the form factor contains information about the longitudinal profile of the bunch, one can only acquire its magnitude $|F(\omega)|$ from the measured intensity distribution. Because $F(\omega) = |F(\omega)| \exp(i\phi(\omega))$, the phases $\phi(\omega)$ are therefore lost during measurement. Recovery of the longitudinal bunch profile is thus a classical 1D phase retrieval problem.
    
    Reconstruction of a 1D signal given the magnitude of its Fourier transform cannot be done uniquely.\cite{Shechtman15, Beinert16} As a result, multiple 1D phase retrieval methods aiming to provide the best estimate for longitudinal bunch profiles have been reported in the literature.\cite{Bajlekov13, BakkaliTaheri16, Su18} A conventional method to recover lost phase information is to extrapolate the form factor's domain to complex frequencies and to assume that this function is analytic. The Kramers-Kronig  (KK) dispersion relation can then be used to derive an expression for the minimal, or KK phases 
    
    \begin{equation} \label{eq:kk_phases}
    	\phi_{KK} (\omega) = -\frac{2\omega}{\pi} \pv{\int_{0}^{\infty} \frac{\ln|F(x)| - \ln|F(\omega)|}{x^2 - \omega^2} \dd{x}} 
    \end{equation}

	 \noindent in terms of the form factor magnitude.\cite{Lai94} Equation (\ref{eq:kk_phases}) represents the lost phases $\phi(\omega)$ only when the complex form factor lacks any zeros in the upper half plane.\cite{Lai95} For general longitudinal bunch profiles found in experiment, this assumption cannot be faithfully made or experimentally verified: a specific case of the non-uniqueness in 1D phase retrieval. 
	
	An alternative approach uses the Gerchberg-Saxton (GS) iterative reconstruction algorithm which converges onto phases that correspond to a real and positive longitudinal profile in the time domain when given a form factor magnitude and array of initial phases.\cite{Gerchberg72, Fienup82} However, different initial phases converge to different reconstructed phase solutions. There is no way of determining whether these reconstructed phases bear close resemblance to the lost phases $\phi(\omega)$, so each bunch profile created using the GS algorithm is a valid solution. 
    
    Historically, the study of reconstruction techniques has focused on the reconstruction of specific test profiles. These profiles are typically well-defined functions like Gaussians, Lorentzians, triangular, or their sum separated by a time interval. However, the results of these studies are limited in their ability to extrapolate findings to arbitrary profiles which can be expected in experiment. Unlike previous studies, we generate a large number of smooth and random longitudinal bunch profiles to assess the accuracy of common bunch reconstruction techniques. 
    
    The paper is structured as follows. In Sec. \ref{sec:GS_algorithm}, we provide an overview of a modified form of the GS algorithm, clarify ambiguities inherent in 1D phase retrieval, and describe a full iteration of the GS algorithm. We then detail an efficient method to mitigate the time shift ambiguity in GS reconstructions, define the average GS reconstruction, and introduce a measure that quantifies uncertainty in GS reconstructions. In Sec. \ref{sec:numerical_experiments}, we describe the creation of smooth and random longitudinal bunch profiles and detail specific methods of longitudinal bunch reconstruction. We then perform numerical experiments employing these reconstruction techniques and draw statistical conclusions about the most appropriate method to reconstruct longitudinal bunch profiles from spectroscopic data.
    
    \begin{figure*}
    	\centering
    	\includegraphics[]{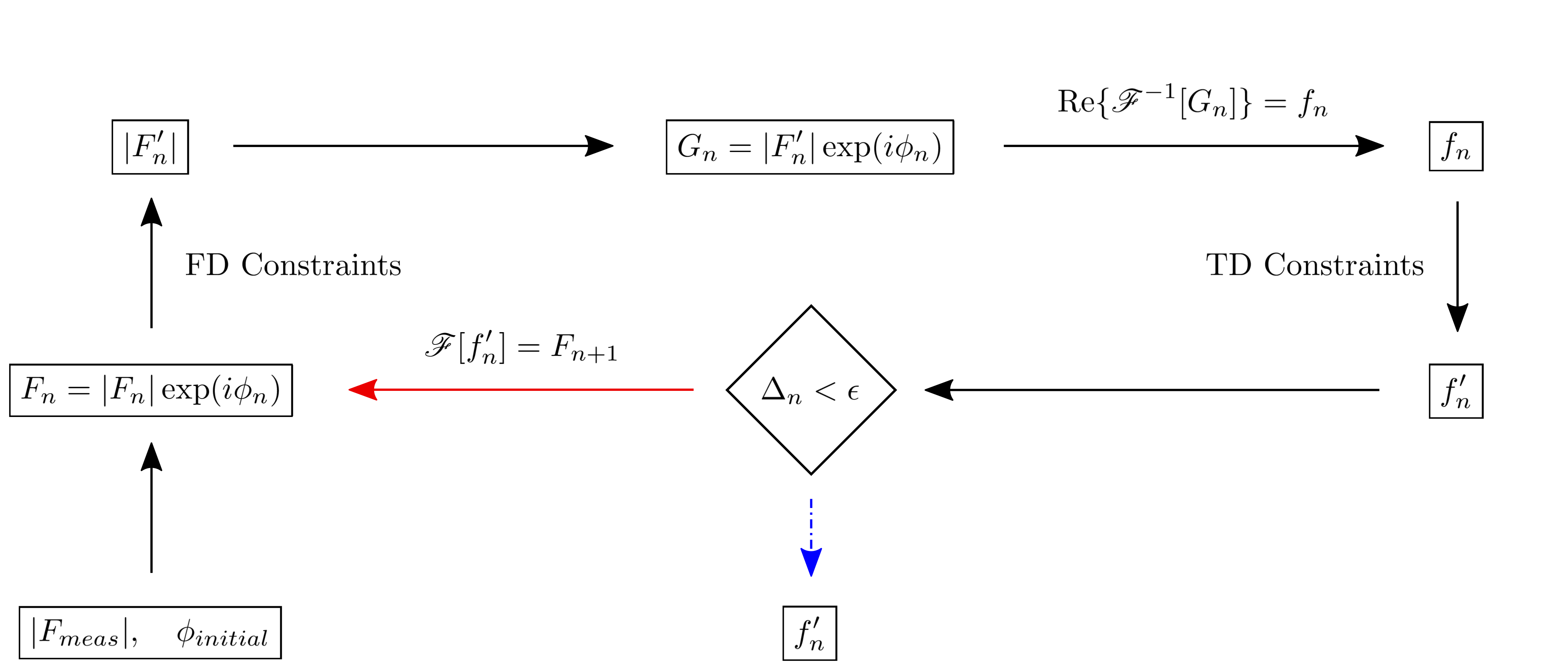}
    	\caption{Overview of the GS algorithm. The measured form factor magnitude $|F_{meas}|$ and initial phases $\phi_{initial}$ are passed in to begin the iterative cycle. The profile $f_n'$ is returned (dotted blue arrow) once the convergence criterion is satisfied, otherwise (solid red arrow) the algorithm continues.}
    	\label{fig:GS_algorithm}
    \end{figure*}
    
    \section{Gerchberg-Saxton Algorithm} \label{sec:GS_algorithm}
    
    The GS algorithm is an iterative procedure to reconstruct Fourier transform phases from a signal's Fourier transform magnitude, first proposed in Ref. \onlinecite{Gerchberg72}. Once the Fourier transform magnitude and initial phases are supplied, the algorithm iteratively applies time and frequency domain constraints to converge onto reconstructed phases approximating the lost phases $\phi(\omega)$. 
    
    Reconstruction of a 1D signal given the magnitude of its Fourier transform is known to be an ill-posed problem. For example, a time shift in the signal or the signal's time reversal results in different Fourier transform phases but an equivalent Fourier transform magnitude. Moreover, multiple non-trivially different signals with different Fourier transform phases can produce an identical Fourier transform magnitude. Although the time shift ambiguity can be ameliorated by means of cross-correlation analysis, reconstructed phases from the GS algorithm still produce time-reversed profiles and profiles with non-trivial ambiguities.
    
    \subsection{Overview of the algorithm}
    
    An overview of the GS algorithm is shown in Fig. \ref{fig:GS_algorithm} which outlines one iteration step. First, the form factor $F_n$ is computed by Fourier transform of the time domain profile $f_{n-1}'$ obtained from the previous iteration step. The magnitude of this form factor is replaced by the form factor magnitude obtained in experiment through the frequency domain (FD) constraint
    
    \begin{equation} \label{eq:fd_constraint}
    	|F_n' (\omega_j)| = |F_{meas} (\omega_j)| .
    \end{equation}
    
    A new form factor which has the phases of $F_n$ and the magnitude $|F_n'|$ is then computed. This form factor is transformed into the time domain through the inverse Fourier transform. The obtained bunch profile is then forced to be strictly positive through the time domain (TD) constraint
    
    \begin{equation} \label{eq:td_constraint}
    	f_n' (t_j) = \abs{f_n (t_j)} .
    \end{equation}
    
    This updated time domain profile serves as a basis for the next iteration step. The algorithm is initiated by providing the form factor magnitude obtained from measured data and an array of initial (typically random) phases. Convergence of the GS algorithm to a specific reconstructed bunch profile is quantified by the correlation coefficient
    
    \begin{equation} \label{eq:correlation_coefficient}
    	r\bqty{g,h}=\frac{\sum{g(t_j)h(t_j)}}   {\sqrt{\sum{g(t_j)^2} \sum{h(t_j)^2}}} ,
    \end{equation}
    
    \noindent where $g$ and $h$ are real-valued functions. If the condition
    
    \begin{equation}
    	\Delta_n \equiv \bigg| r \big[ \abs{F_{n+1}}, \abs{F_{meas}} \big] - r \big[ \abs{F_{n}}, \abs{F_{meas}} \big] \bigg| < \epsilon
    \end{equation}
    
    \noindent is satisfied for a convergence parameter $\epsilon$, typically on the order of $10^{-9}$, then the GS algorithm returns the reconstructed profile $f_{n}'$ (dotted blue arrow in Fig. \ref{fig:GS_algorithm}). If not, the algorithm begins another iteration (solid red arrow in Fig. \ref{fig:GS_algorithm}). 
    
    We define $\Delta_n$ to be the absolute difference in correlation coefficients between subsequent iteration steps because it allows for the successful reconstruction of many longitudinal profiles using a single value of $\epsilon$ regardless of the initial phases chosen. Once $\Delta_n$ becomes smaller than a well-chosen $\epsilon$, subsequent iterations do little to improve the correlation between $\abs{F_n}$ and $\abs{F_{meas}}$ (and therefore do not create any perceptible improvement in the reconstruction $f'_n$). Convergence of the GS algorithm is defined to be at this point. 
    
    \subsection{Modifications to the algorithm}
    
    Because individual reconstructions returned by the GS algorithm are not unique, we can create many reconstructions and average them to approximate the longitudinal bunch profile as first proposed in Refs. \onlinecite{Pelliccia12, Pelliccia14}. No individual reconstruction is discarded since every array of converged phases, and hence every GS reconstruction, is experimentally valid. Random initial phases are uniformly chosen from the interval $\lbrack 0,2\pi)$ when creating individual reconstructions to ensure that each outcome of the GS algorithm is equally likely to occur. Features present in the majority of reconstructed profiles will be prominent in the average profile and random fluctuations unique to an individual reconstruction will be averaged out. 
    
    For this procedure to work, one must impose a time shift to each individual reconstruction, otherwise the average reconstruction will simply be a uniform background profile. Similarly, one must decide whether an individual reconstruction should be time-reversed, otherwise the average reconstruction will be symmetric. To perform these shifts and potential time reversals in a self-consistent manner, one can algorithmically shift and (or) flip an individual reconstruction $f$ so as to maximize its cross-correlation
    
    \begin{equation} \label{eq:cross_correlation_eq}
    	c(\tau) = \int_{-\infty}^{\infty} f(t) g(t+\tau) \dd{t}
    \end{equation}
    
    \noindent with a reference profile $g$. The choice of reference profile can be any reconstruction produced by the GS routine; since we assume no further knowledge about the longitudinal bunch profile $f^0$ to determine a judicious choice beyond its form factor magnitude, we choose the first reconstruction $f_{GS}^1$  for simplicity. A straightforward but inefficient approach first computes $c(\tau)$ $N_t$ times where $N_t$ is the number of discrete points in the time domain describing $f$ and $g$. This brute-force computation is then done $N_t$ times again for $f(-t)$, after which one finally determines the shift $\tau$ and potential time-reversal for $f$ that maximizes its cross-correlation with $g$.
    
    Alternatively, we use an efficient approach to maximize the cross-correlation between $f$ and $g$. By the Cross-Correlation Theorem, $\mathscr{F}\bqty{c(\tau)} = F(\omega) G^* (\omega)$ where $F(\omega) = \mathscr{F} \bqty{f}$ and $G(\omega) = \mathscr{F} \bqty{g}$. Hence
    
    \begin{equation}
    	c(\tau) = \frac{1}{2\pi} \int_{-\infty}^{\infty} F(\omega) G^* (\omega) e^{i\omega \tau} \dd{\omega} .
    \end{equation}
    
    \noindent The cross-correlation is maximized when
    
    \begin{equation}
    	\dv{c}{\tau} = \frac{1}{2\pi}\int_{-\infty}^{\infty} i \omega F(\omega) G^* (\omega) e^{i\omega \tau} \dd{\omega} = 0 .
    \end{equation}
    
    Defining $h(\tau) = \Re \Bqty{\mathscr{F}^{-1} \bqty{i \omega F(\omega) G^* (\omega)}}$, solving for $h(\tau) = 0$ yields a time shift $\tau$ that maximizes the cross-correlation $c(\tau)$. The computational complexity of this procedure is $\mathcal{O} (N_t \log N_t)$ as opposed to $\mathcal{O}(N_t^2)$ through brute force. 
    
    The procedure is repeated for $f(-t)$ to account for time-reversal ambiguity, after which the shifted (and potentially time-reversed) profile that best correlates with $g$ is selected. The shifts found through this efficient method and through brute-force are identical. Approximately $50\%$ of reconstructions are flipped when a sufficiently large number of reconstructions are created. This is precisely the percentage of flipped solutions to be expected since individual reconstructions are randomly oriented due to the previously mentioned time-reversal ambiguity. After each GS reconstruction is best correlated with the reference profile $g$, the average reconstruction and rms deviation from the average reconstruction (similar to Refs. \onlinecite{Bajlekov13, Hass18}) are computed:
    
    \begin{eqnarray} 
    	\avg{f_{GS}(t_j)} &=& \frac{1}{N} \sum_{i=1}^{N} f_{GS}^i (t_j) \label{eq:avg_reconstruction}\\
    	\sigma (t_j) &=& \sqrt{\frac{1}{N} \sum_{i=1}^{N} \big( f_{GS}^i (t_j) - \avg{f_{GS}(t_j)} \big)^2} \label{eq:rms_deviation}.
    \end{eqnarray}
    
    Here, $N$ is the number of reconstructions created with the GS algorithm and $f_{GS}^{i}$ is the $i\textsuperscript{th}$ GS reconstruction (created using random initial phases). The rms deviation $\sigma(t_j)$ can be viewed as an uncertainty in the algorithm used to recover $f^0$, similar to the error obtained in experimental measurements, and is a useful and self-consistent statistical measure to quantify the difference in non-unique GS reconstructions.
    
    \section{Numerical Experiments} \label{sec:numerical_experiments}
    
    \subsection{Longitudinal profile generation}
    
\begin{figure}
		\centering
		\includegraphics[]{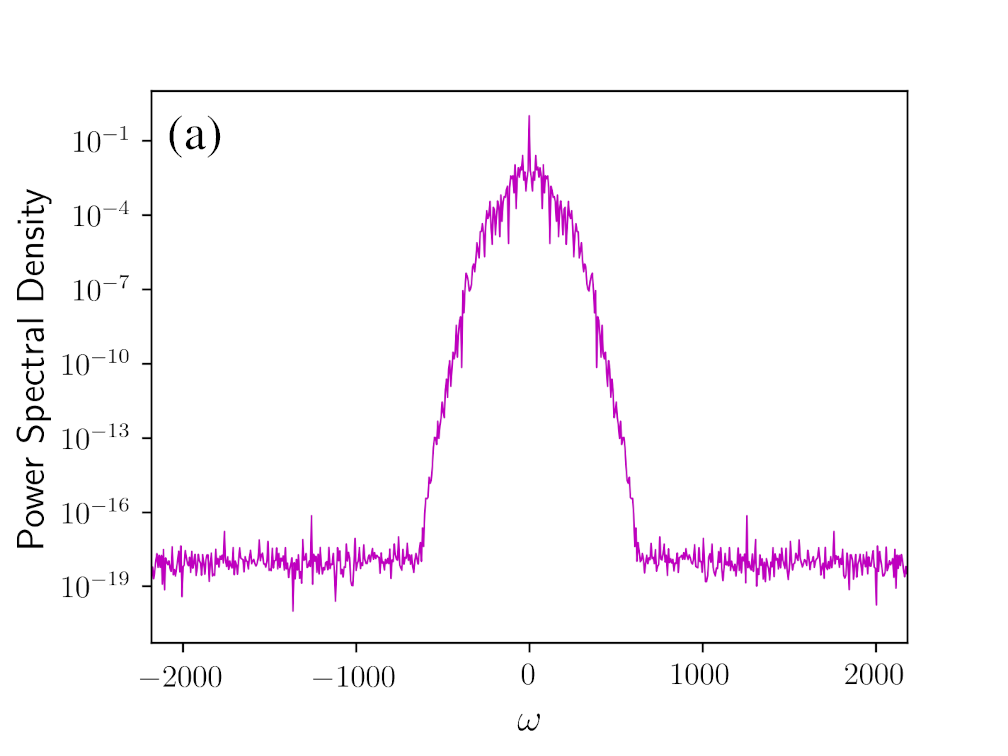}
		\includegraphics[]{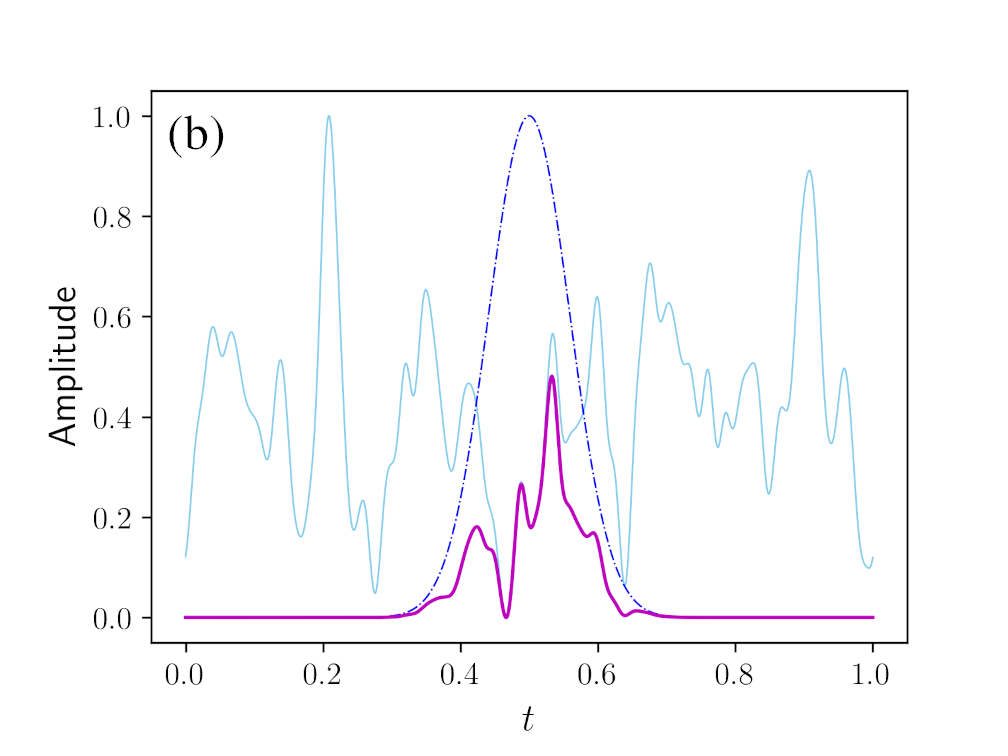}
		\caption{(a) Power spectral density of a smooth and random series with $N_{var}=6.0$. (b) A randomly generated longitudinal profile with $N_{var} = 6.0$. Depicted are the smooth and random series (solid light blue), Gaussian envelope (dashed blue), and unnormalized longitudinal profile (solid magenta).}
		\label{fig:random_profile_generation}
	\end{figure}

    To generate smooth and random periodic profiles, we first create a series of random and uncorrelated values on a time domain grid. A $7$-point smoothing average is applied to this series so that neighboring grid points are correlated. Smoothing increases the rms length of the autocorrelation function for this series. Additionally, the series becomes periodic if the smoothing involves circular shifts. A smoothing average is applied in circular shifts multiple times until the desired rms correlation length is reached. The resulting series approximates a stationary process with a power spectral density that resembles a Gaussian distribution. The power spectral density of a smooth and random series generated in this manner is presented in Fig. \ref{fig:random_profile_generation}a.
	
	The resulting series is smooth and periodic (to avoid artificially large amplitudes of high harmonics in its spectrum). However, we aim to reconstruct isolated longitudinal profiles. For that reason, we multiply the smooth and random series by a Gaussian envelope. The degree of fluctuation in the resulting longitudinal profile is defined by the parameter $N_{var}$, which equals the ratio of Gaussian rms length to rms length in the autocorrelation function of the smooth and random series. Large values of $N_{var}$  correspond to longitudinal profiles with a larger number of peaks and valleys under the Gaussian envelope. Small values of $N_{var}\lesssim1$ correspond to approximately Gaussian longitudinal profiles. An example longitudinal profile created through this process is shown in Fig. \ref{fig:random_profile_generation}b.
	
	This procedure for generating smooth and random longitudinal profiles with identical statistical properties (\textit{i.e.} $N_{var}$) allows for systematic studies on the accuracy of several reconstruction methods.
	
	\subsection{Reconstruction methods} \label{sec:reconstruction_methods}
	
	In the following numerical experiments, we use four methods to create reconstructions using a form factor magnitude $\abs{F}$ corresponding to a randomly generated longitudinal bunch profile:
	
	\begin{enumerate}
		\item Equation (\ref{eq:kk_phases}) is numerically evaluated using $\abs{F}$. The form factor $\abs{F}\exp(i\phi_{KK})$ is then inverse Fourier transformed to create the longitudinal profile $f_{KK}$.
		\item The KK phases and form factor magnitude $\abs{F}$ are input into the GS algorithm until convergence, leading to the profile $f_{KK-GS}$ as proposed in Ref. \onlinecite{Schmidt20}. This reconstruction's form factor magnitude is unchanged while negativity in the time domain, common in reconstructions using KK phases, is removed.
		\item The GS algorithm is employed using $\abs{F}$ and random initial phases to create the reconstruction $f_{GS}$.
		\item Multiple GS reconstructions, each corresponding to different initial conditions, are shifted and (or) flipped through cross-correlation analysis with the reference profile $f_{GS}^1$ and then averaged using Eq. (\ref{eq:avg_reconstruction}) to create the average reconstruction $\avg{f_{GS}}$.
	\end{enumerate}
	
	The accuracy of each reconstruction method is quantified by the correlation coefficient $r$, given by Eq. (\ref{eq:correlation_coefficient}). It represents the normalized $L^2$ inner product between two functions and indicates how similar these functions are. Identical functions result in $r=1$.
	
	\begin{figure*}
		\centering
		
		\begin{subfloat}
			\centering
			\includegraphics[]{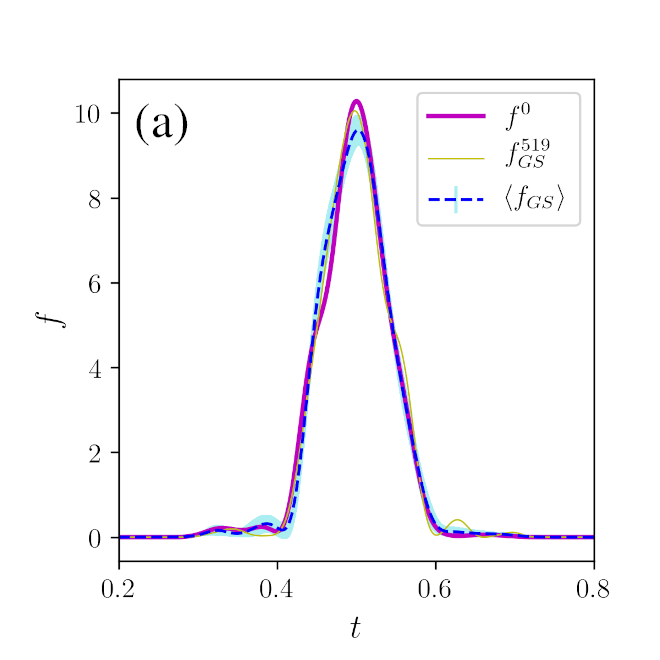}
		\end{subfloat}
		\begin{subfloat}
			\centering
			\includegraphics[]{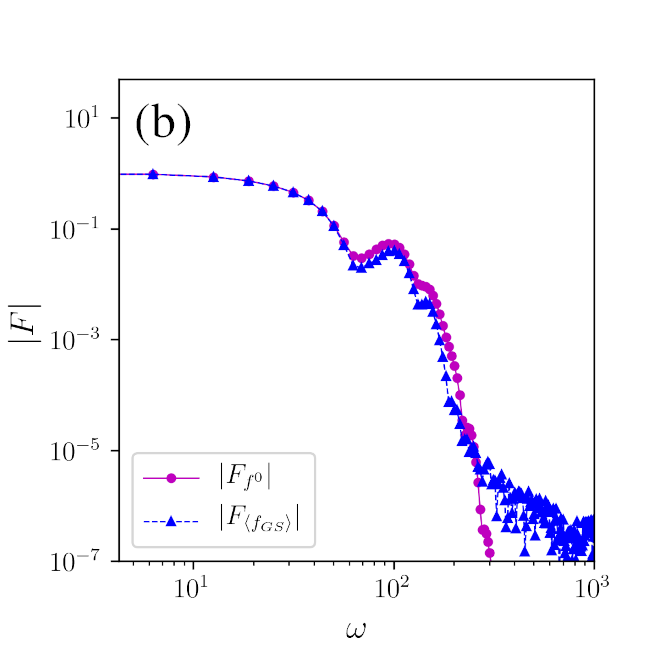}
		\end{subfloat}
		\begin{subfloat}
			\centering
			\includegraphics[]{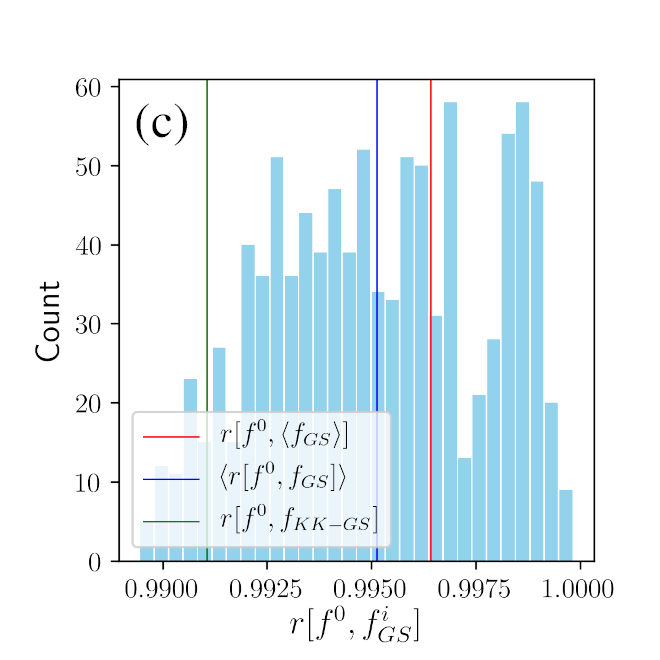}
		\end{subfloat}
		
		\begin{subfloat}
			\centering
			\includegraphics[]{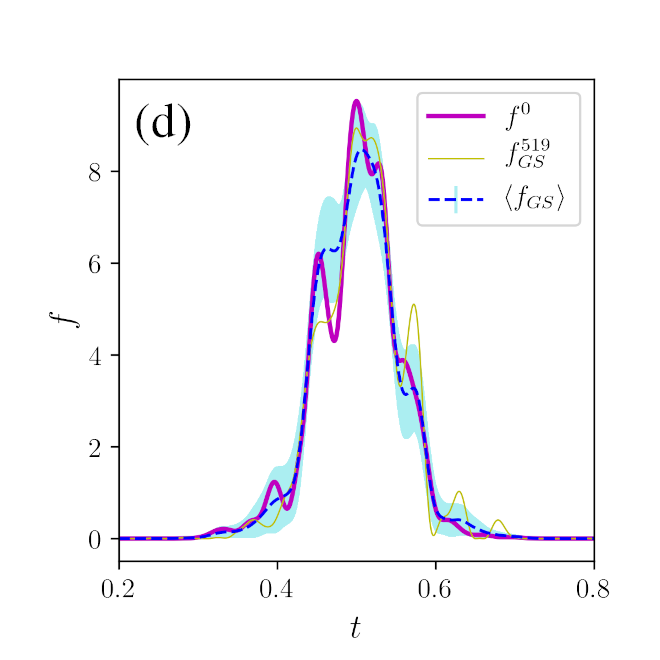}
		\end{subfloat}
		\begin{subfloat}
			\centering
			\includegraphics[]{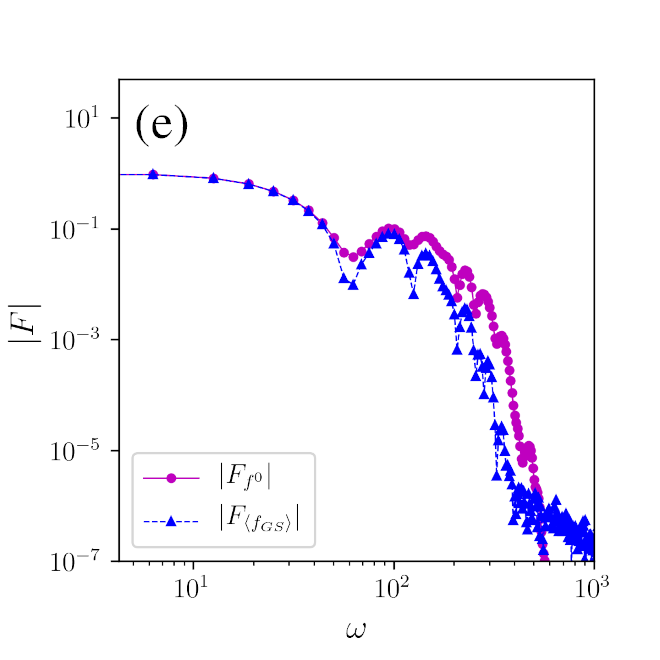}
		\end{subfloat}
		\begin{subfloat}
			\centering
			\includegraphics[]{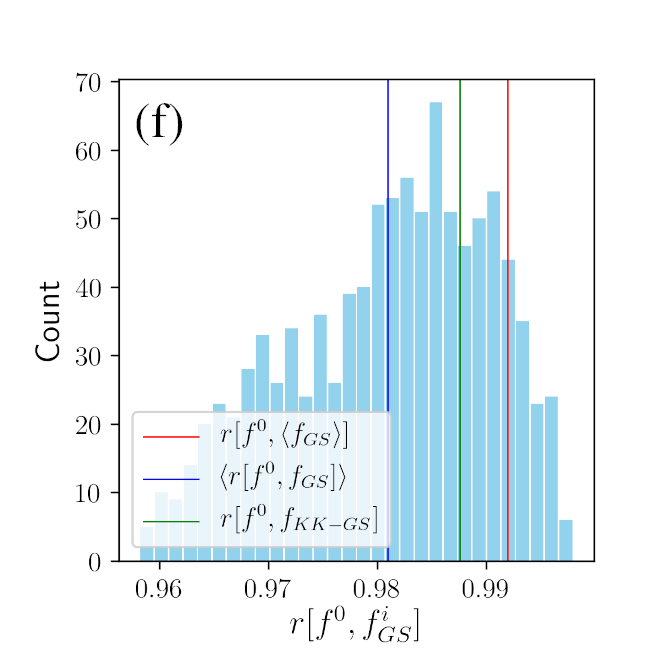}
		\end{subfloat}
		
		\caption{(a), (d): Plots of a randomly generated longitudinal profile $f^0$, randomly selected individual GS reconstruction $f_{GS}^{519}$, reconstruction $f_{KK-GS}$ obtained from the GS algorithm seeded with $\phi_{initial} = \phi_{KK}$, and average reconstruction $\avg{f_{GS}}$ created from $1000$ individual GS reconstructions. Plot (a) shows $f^0$ with small variability ($N_{var} = 3$) in contrast to plot (d) with large variability ($N_{var}=6$). (b), (e): Form factor magnitudes for $f^0$ and $\avg{f_{GS}}$ given in plots (a) and (d), respectively. (c), (f): Correlation coefficient histograms comprised of $1000$ individual GS reconstructions.}
		
		\label{fig:noisefree_profiles}
	\end{figure*}
	
	\subsection{Reconstruction of noise-free profiles} \label{sec:noise_free_reconstruction}
	
	Plots (a) and (d) in Fig. \ref{fig:noisefree_profiles} each show a randomly generated longitudinal bunch profile, denoted as $f^0$. The profile $f^0$ in plot (a) has a small amount of fluctuation ($N_{var}=3$), while $f^0$ in plot (d) contains more variability ($N_{var} = 6$). Furthermore, each plot contains three reconstructions: a randomly selected GS reconstruction $f_{GS}^{519}$ (out of $1000$ total GS reconstructions created in the study), the KK-GS reconstruction, and the average reconstruction $\avg{f_{GS}}$. Blue shaded regions denote the rms deviation from the average reconstruction, given by Eq. (\ref{eq:rms_deviation}). Plots (b) and (e) demonstrate how the form factor magnitude of the average reconstruction $\abs{\mathscr{F}\bqty{\avg{f_{GS}}}}$ differs from the form factor magnitude of $f^0$, and plots (c) and (f) show the accuracy of each reconstruction technique through correlation coefficient histograms. 
	
	Several results from these two simulations are of note. First, they clearly demonstrate that 1D phase retrieval, whether through the KK phase relationship or GS algorithm, is non-unique. These reconstructions may under- or over-estimate the peaks and valleys in $f^0$. Plots (c) and (f) demonstrate that the average reconstruction $\avg{f_{GS}}$ correlates with the randomly generated longitudinal profile higher than any single reconstruction, whether from the GS algorithm or by means of the KK phase relationship. These plots also show that individual GS reconstructions have a large spread in their correlation coefficients with $f^0$. 
	
	Individual GS reconstructions converge to solutions which have form factor magnitudes matching that found through observation. Averaging solutions in the time domain is equivalent to averaging their form factors in the frequency domain. Unless these form factors have identical phases, this averaging results in a form factor with different magnitude than individual form factors. As a result, the average reconstruction $\avg{f_{GS}}$ has a form factor magnitude $\abs{\mathscr{F}\bqty{\avg{f_{GS}}}}$ which differs from the measured form factor magnitude $\abs{\mathscr{F}\bqty{f^0}}$.
	
\begin{figure}
		\centering
		\includegraphics[]{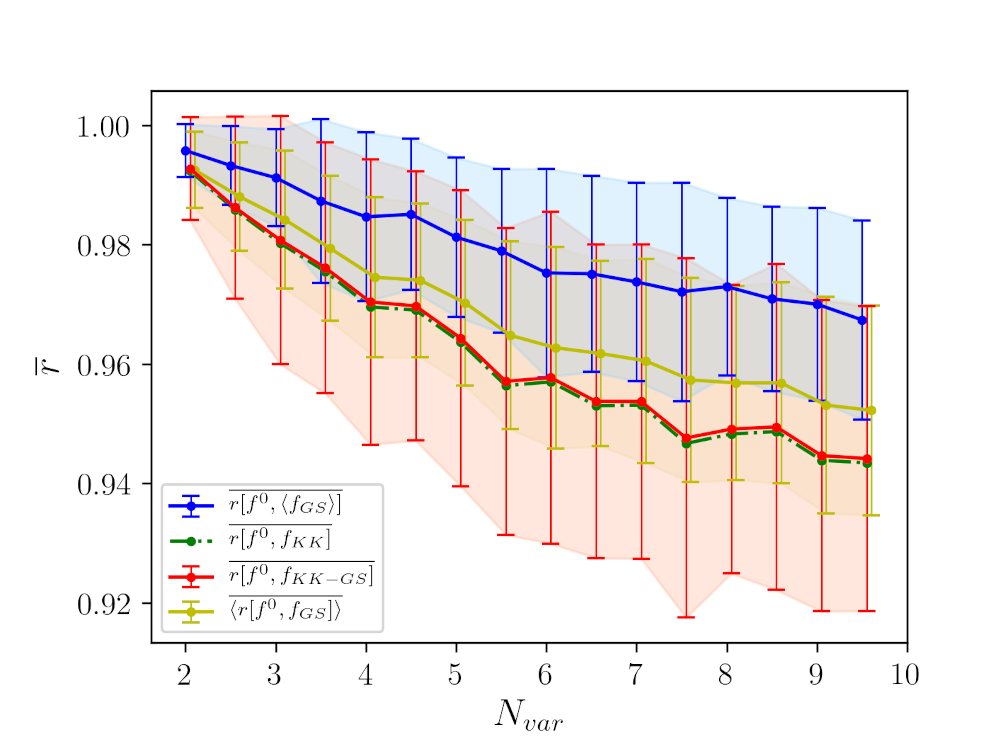}
		\caption{Correlation coefficients, averaged over $200$ runs, between randomly generated longitudinal profiles with variation $N_{var}$ and KK, KK-GS, individual GS, and average GS reconstructions. Graphs corresponding to different methods have small longitudinal offsets relative to each other for better visual representation. KK and KK-GS rms deviations are similar, so the KK rms deviation is omitted.}
		\label{fig:corr_vs_nvar}
	\end{figure}

	To determine the most accurate reconstruction method of those introduced for a wide range of longitudinal bunch profiles, the following numerical experiment was implemented. First, a random longitudinal profile with a given $N_{var}$ was generated. The KK reconstruction, KK-GS reconstruction, 250 GS reconstructions, and average GS reconstruction were created as described in Sec. \ref{sec:reconstruction_methods} and their correlation coefficients with the randomly generated longitudinal profile were stored. Additionally, the correlation coefficient between each of the $250$ individual GS reconstructions and the randomly generated longitudinal profile was calculated and the average of these correlation coefficients (denoted as $\avg{r\bqty{f^0, f_{GS}}}$) was stored. The respective correlation coefficients were averaged over $200$ runs (this averaging is denoted by an overhead bar in Fig. \ref{fig:corr_vs_nvar}). 
	
	Figure \ref{fig:corr_vs_nvar} demonstrates that the average GS reconstruction (blue graph) approximates original longitudinal profiles statistically better than all other studied reconstruction methods. All averaged correlation coefficients decrease with increased variability in the longitudinal profile, but the average GS reconstruction results in the most favorable scaling in averaged correlation coefficients. Reconstructions using KK phases perform the worst, while individual GS reconstructions using randomly chosen initial phases fair slightly better. Moreover, KK and KK-GS reconstructions have a larger correlation coefficient spread than individual GS reconstructions. From these results, we conclude that averaging several shifted and (or) flipped GS reconstructions (created using different random initial phases) produces, on average, a more accurate reconstruction of  longitudinal bunch profiles than one would obtain through the KK, KK-GS, or individual GS reconstruction techniques. 
	
	\begin{figure}
		\centering
		\includegraphics[]{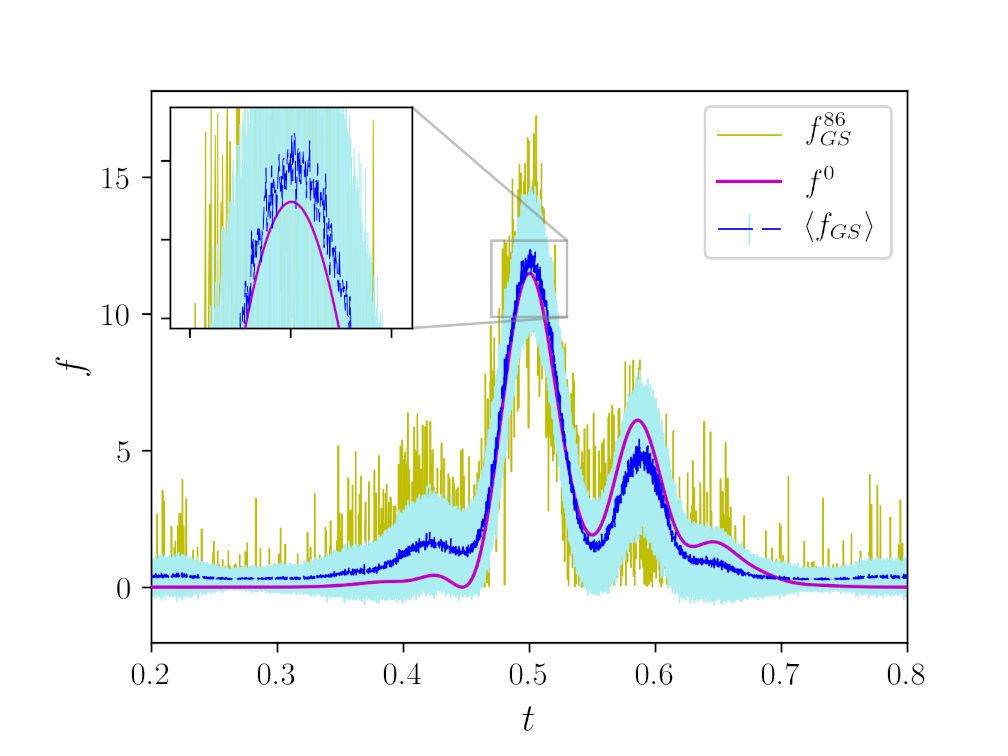}
		\caption{Reconstruction of a randomly generated $N_{var}=3.0$ profile from its noisy form factor magnitude with a noise level of $\Delta_f = 0.018$. $1000$ individual noisy GS reconstructions were used to create $\avg{f_{GS}}$. The light-blue shaded region depicts the rms deviation from the noisy average reconstruction.}
		\label{fig:noisy_profiles}
	\end{figure}
	
    \subsection{Reconstruction of profiles from noisy data}
    
    The noise present in real intensity distribution data (and hence measured form factor magnitudes) must be taken into account when determining the appropriate longitudinal bunch reconstruction technique. We introduce the noise function $n(\omega_j) = \max \{ \abs{F} \} \times X$ where $X \sim \mathcal{N}(0,\Delta_f^2)$ is a Gaussian-distributed random variable. The parameter $\Delta_f$ serves as a quantitative measure of noise level for this study. The form factor magnitude
    
    \begin{equation}
    	\abs{G (\omega_j)} = \abs{F (\omega_j) + n(\omega_j) \exp(i \phi (\omega_j))}
    \end{equation}
    
    \noindent is used as a noisy input for the reconstruction methods detailed in Sec. \ref{sec:reconstruction_methods}. Each noise phase $\phi(\omega_j)$ is randomly generated from a uniform distribution over $\lbrack 0, 2\pi)$. Figure \ref{fig:noisy_profiles} shows the reconstruction of a randomly generated profile from its noisy form factor magnitude created through this method. 
    
\begin{figure}
		\centering
		\includegraphics[]{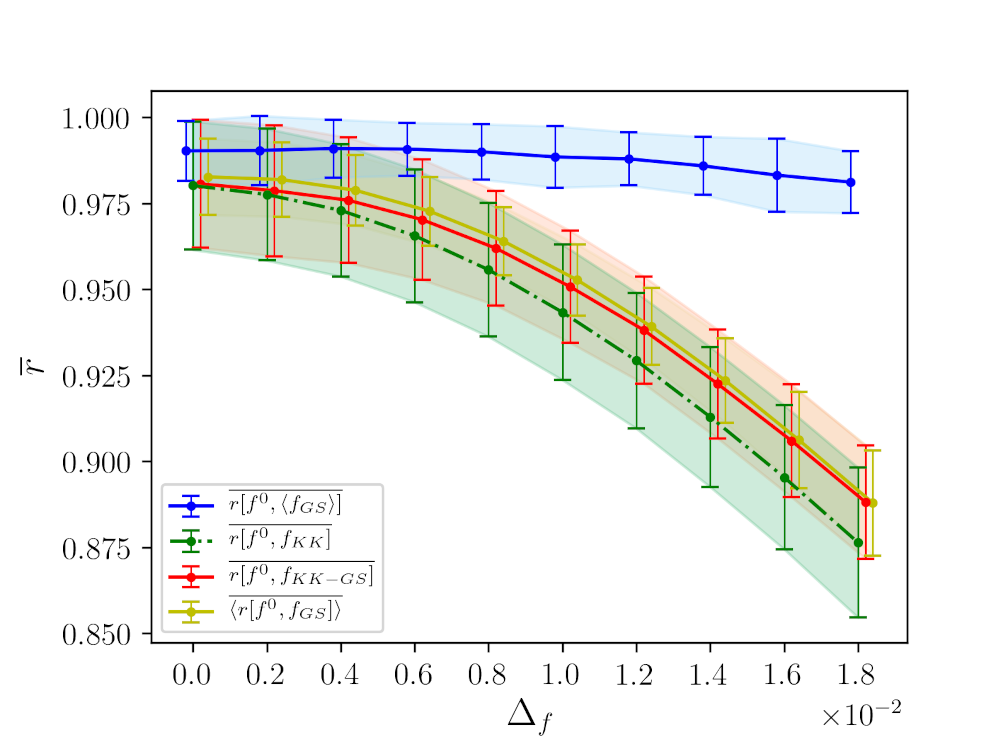}
		\caption{Accuracy of various reconstruction algorithms as a function of noise level introduced in form factor magnitudes. Algorithms were tested for longitudinal profiles with $N_{var}=3.0$.}
		\label{fig:corr_vs_errorfrac}
	\end{figure}

    In the same manner as Sec. \ref{sec:noise_free_reconstruction}, we reconstructed $200$ different longitudinal profiles having $N_{var} = 3.0$ and a given value of $\Delta_f$. The number of reconstructions is sufficient to accumulate enough statistics to quantify the accuracy of each reconstruction algorithm. The KK, KK-GS, individual GS, and average GS reconstructions were created for each $\Delta_f$ and their respective correlation coefficients with the noiseless longitudinal profile $f^0$ were stored. These correlation coefficients were then averaged over all runs and are illustrated in Fig. \ref{fig:corr_vs_errorfrac}.
    
    The results show that the average GS reconstruction performs much better than other elucidated reconstruction techniques when noise is present in the form factor magnitude. The averaged correlation coefficient graph for the average GS reconstruction is mostly flat over a large range of noise amplitudes, while averaged correlation coefficients from other reconstruction methods sharply decrease in accuracy with increased amounts of noise. Notably, the KK-GS and individual GS reconstruction graphs perform on-par with one another for larger noise levels, while the KK graph performs the worst due to negativity in the time domain of KK reconstructions. 
    
    The KK, GS, and KK-GS methods produce individual noisy longitudinal profiles when the form factor magnitude used for reconstruction contains noise. On the other hand, reconstruction based on averaging many GS reconstructions preserves features common to individual reconstructions and washes out irregularities. This is the reason for the average GS reconstruction's superior accuracy compared to other reconstruction methods when considering noisy spectrum data.
	
    \section{Summary}
    
    In this work, we have studied different techniques for reconstructing electron bunch profiles in the time domain from form factor magnitude (power spectrum) data. In our study, we focused on quantifying the accuracy of different reconstruction techniques in a self-consistent manner. To address this goal, we generated multiple smooth and random longitudinal profiles with identical statistical properties and attempted to reconstruct these profiles from their form factor magnitudes. A large number of random longitudinal profiles allows for statistical analysis of each reconstruction method and quantitative comparison between them.
    
    We have evaluated the accuracy of reconstruction techniques employing the Kramers-Kronig phase relationship, Gerchberg-Saxton algorithm, their combination, and an algorithm based on averaging many individual GS reconstructions. The algorithm averaging many individual GS reconstructions significantly outperforms other studied algorithms, particularly in scenarios involving large amounts of noise in measurements. This algorithm is a superior candidate for real 1D electron bunch profile diagnostics based on spectral measurements. 
    
    \section*{Acknowledgments}
    
    We gratefully acknowledge the support of the US Department of Energy through the LANL LDRD program for this work. Los Alamos National Laboratory is operated by Triad National Security, LLC. 
    
    \section*{Data Availability}
    
    The data that support the findings of this study are available from the corresponding author upon reasonable request.
    
    \bibliography{bibliography}

\end{document}